\documentclass{npcs}
\usepackage{graphics} 
\usepackage{epsfig}
\usepackage{amsmath, amssymb}
\begin{document}
\topmargin = -30mm

\runauthor{V. Makarenko, T. Shishkina}
\runtitle{
The finite mass calculations for $\gamma \gamma \to l^{+} l^{-} \gamma$ process
}

\begin{topmatter}
\title{
The finite mass calculations for $\gamma \gamma \to l^{+} l^{-} \gamma$ process
}
\author{Author V.V. Makarenko}
\institution{NC PHEP BSU}
\email{makarenko@hep.by}
\author{Author T.V. Shishkina}
\institution{NC PHEP BSU}
\email{shishkina@hep.by}
\vspace{1cm}
\end{topmatter}

\mathchardef\vm="117
\mathchardef\um="11D
\mathchardef\E="245
\mathchardef\Mom="250
\def\z1{z{}'}
\def\t1{t{}'}
\def\u1{u{}'}
\def\s1{s{}'}
\def\m2{m^2}
\def\d12{\frac{1}{2}}
\def\lfr#1#2{\ln{\frac{#1}{#2}}}

\def\ub{\bar{\um}}
\def\Jint#1{\mathcal{J} ( {#1} )}
\def\Iint#1#2{\int\limits_{0}^{\ub}{\Jint{#1}} {#2} {d \um}}
\def\Iintl#1#2{\int\limits_{2 m \lambda}^{\ub}{\Jint{#1}} {#2} {d \um}}

\def\bracket#1{\left({#1}\right)}
\def\bra#1{\bracket{#1}}
\def\spr#1#2{\, #1 \! \cdot \! #2 \,}
\def\ddx#1{\frac{1}{#1}}


The luminosity of photon beams at the future $\gamma\gamma$-colliders \cite{tdr, gg_proposal}
will be measured using the well-known QED reactions of light lepton pairs creation
($\gamma\gamma\to l^{+} l^{-}$ \cite{gg_2f}, $\gamma\gamma\to 4 l$ \cite{gg_4f} and others \cite{tdr, gg_llg_lumi}).
It turned out, that in the experiments
using beams of similar helicity (total helicity of $\gamma\gamma$-system $J\!=\!0$)
the most of these processes are greatly suppressed and can't be used.
However these experiments are extremely significant for the light Higgs boson searching \cite{tdr}.

It was shown \cite{gg_llg_lumi} that the process of bremsstrahlung $\gamma\gamma\to l^{+} l^{-} \gamma$
can be used successfully for the luminosity measurement in the experiments with $J\!=\!0$-beams.

The analysis performed in ref. \cite{gg_llg_lumi}
is based on the method of helicity amplitudes \cite{ha} and the masses of leptons were neglected.
Indeed, the mass of muon (and electron as well) is about $10^4$ times smaller than the experimental energies.
But the energies of final particles in considering kinematics can be small enough (down to $1 GeV$)
when they are compared to muon mass.
The cross section of process at this edge of kinematical regions can also be very large.
It can lead to the uncertainties in calculations (see our previous paper \cite{gg_llg_lumi}).

In this paper the process $\gamma\gamma\to l^{+} l^{-} \gamma$ is analysed
on a tree level for the fixed lepton mass.
The fixed-mass calculations are performed precisely
and the finite lepton mass was kept throughout the analysis.
The results are compared to ref. \cite{gg_llg_lumi}.

We consider the process

$$ \gamma(p_1, \lambda_1)+\gamma(p_2, \lambda_2) \to l(p_1{}', e_1{}') + \bar{l}(p_2{}', e_2{}') + \gamma(k, \lambda_k).$$

The notations are similar to ones used in ref. \cite{gg_llg_lumi}.
We denote the c.m.s. energy squared by $s = {\left(p_1+p_2\right)}^2 = 2 \spr{p_1}{p_2},$
the final-state photon energy by $w$.
For the differential cross-section we introduce the normalized
final-state photon energy (c.m.s. is used) $
x=w\slash\sqrt{s}$.
The differential cross section ${d\sigma}\slash{d x}$
appears to be the energy spectrum of
final-state photons.

Since the final-state polarizations can't be measured we summarize
over all the final particles helicities.
The integration over the phase space of final particles is performed numerically using the
Monte-Carlo method \cite{mc}.

According to \cite{gg_llg_lumi} we introduce the set of
restrictions on
the parameters of final particles.
The events are not detected if energies and angles are below the corresponding threshold values.
We use the following restrictions on the phase-space of final particles \cite{tdr}:

$\bullet$ Minimum final-state photon energy: $\omega_{cut}$,

$\bullet$ Minimum fermions energy: $E_{f,cut}$,

$\bullet$ Minimum angle between any final and any initial particles (polar angle cut): $\Theta_{cut}$,

$\bullet$ Minimum angle between any pair of final particles: $\varphi_{cut}$.

The graphs presented below are similar to ones introduced in paper \cite{gg_llg_lumi},
but calculated using precise formulae.
For each plot we introduce the graph for relative correction $$\delta = \sigma_m \slash \sigma_0 -1.$$

All the plots are composed for process $\gamma\gamma\to\mu\bar{\mu}\gamma$.

It was discovered that the relative corrections for the total cross sections at considering cuts
are at the level of $1\%$.

The corrections to the differential cross sections
are also small
except the single points, where it can rise up to $100\%$.
The high errors occur in the $J\!=\!0$ energy plot near the $w\to 0$ limit
and in some of angular spectra of final particles.

\begin{figure}[h!]
\leavevmode
\begin{minipage}[h]{.5\linewidth}
\centering
\includegraphics[width=\linewidth, height=2.7in, angle=0]{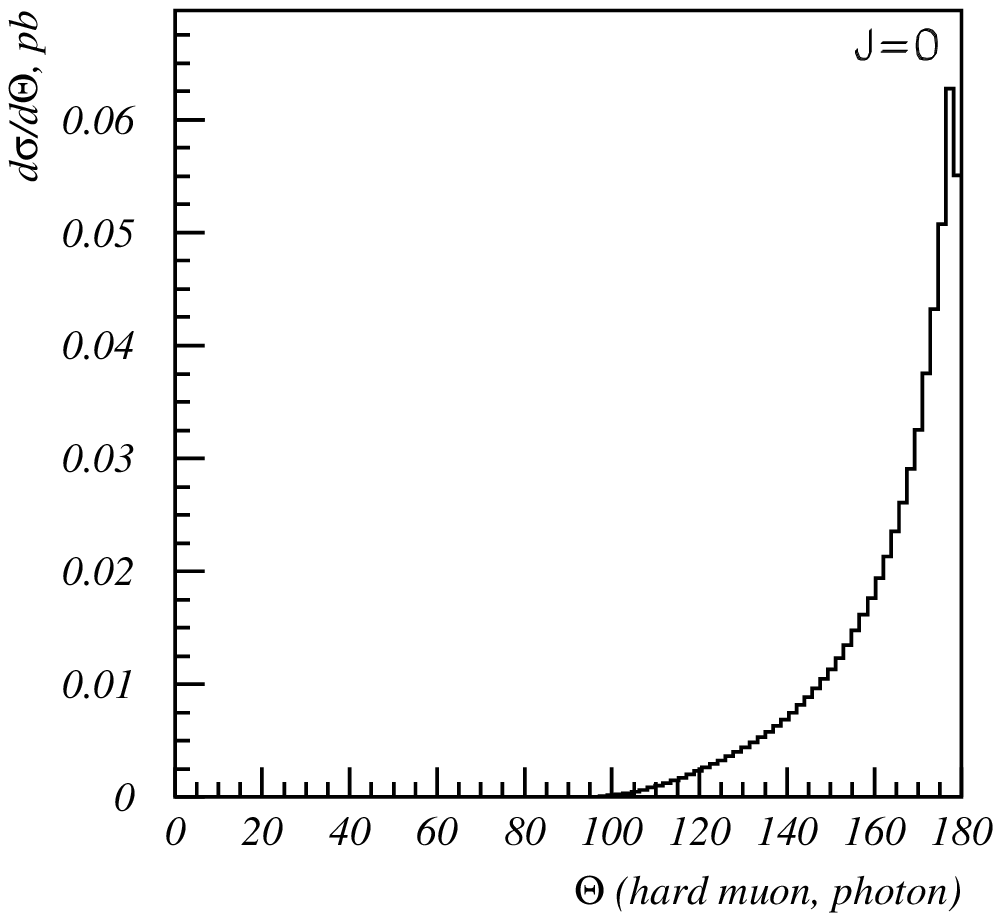}
\end{minipage}\hfill
\begin{minipage}[h]{.5\linewidth}
\centering
\includegraphics[width=\linewidth, height=2.7in, angle=0]{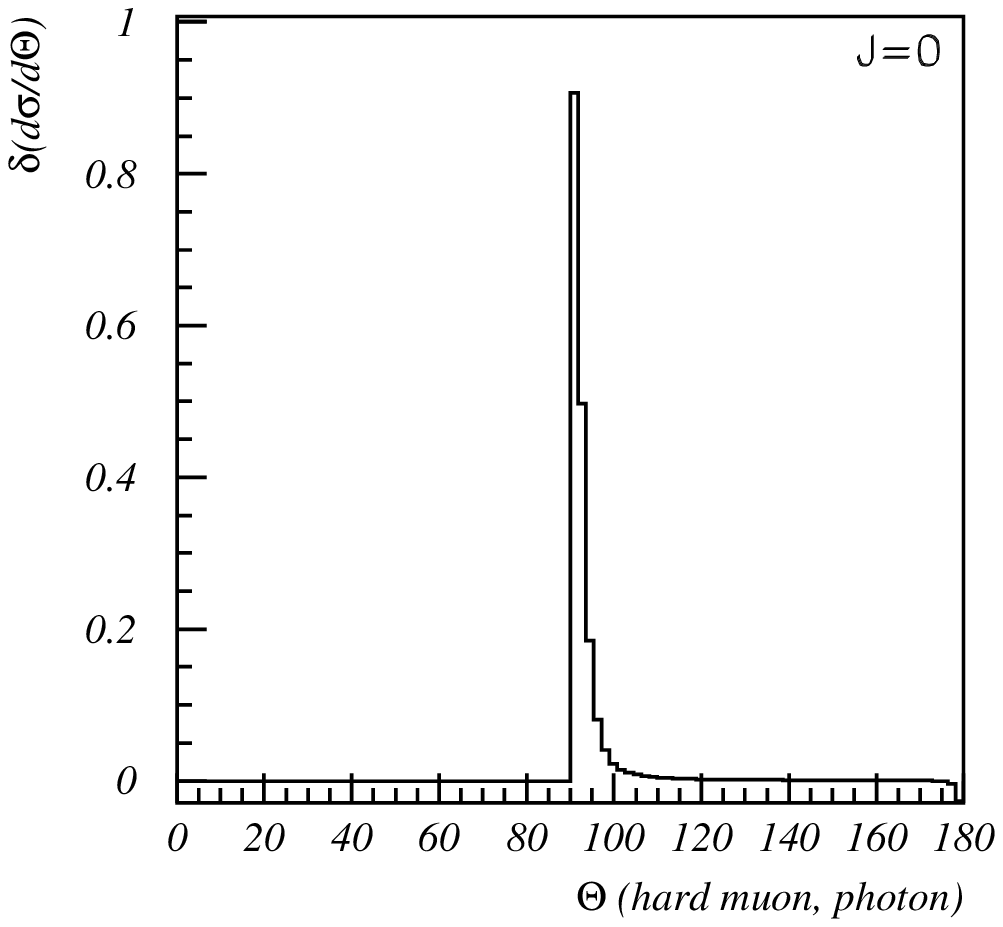}
\end{minipage}
\caption{
Angular spectrum of final particles,
$\sqrt {s} = 120 GeV$,
$w_{cut}\!=\!1GeV$, $\E_{cut}\!=\!1GeV$, $\Theta_{cut}\!=\!7^o$, $\varphi_{cut}\!=\!3^o$.
}\label{f_hard_1}
\end{figure}

Consider the graph \ref{f_hard_1} in details.
It is the plot for the differential cross section
on the angle between the final photon and the hardest muon.
The kinematical restriction on the lower edge of this graph is defined by formula
$$ \varphi_{min}(\gamma, \mu_{hard})=\frac{\pi}{2}+
\arcsin{\bra{\omega_{cut}\slash{\sqrt{{\bra{\E-\omega_{cut}}}^{2}-4 m^{2}}}}}.$$
The higher cut on the final photon energy leads to the higher minimal angle between photon and hard muon.
It means that all the events near the left edge of graph have the energy of final photon close to $\omega_{cut}$.
Hence if the $\omega_{cut}$ is small enough (when compared to muon mass)
the left edge of plot should outbreak the high error in massless calculations.

Due to small value of mass contribution
we do not present any angular spectra for $J\!=\!2$ beams
as well as plots for angle between soft muon and photon for $J\!=\!0$ beams.
The finite muon mass affects significantly only the edges of that plots.

\begin{figure}[h!]
\leavevmode
\begin{minipage}[h]{.5\linewidth}
\centering
\includegraphics[width=\linewidth, height=3.5in, angle=0]{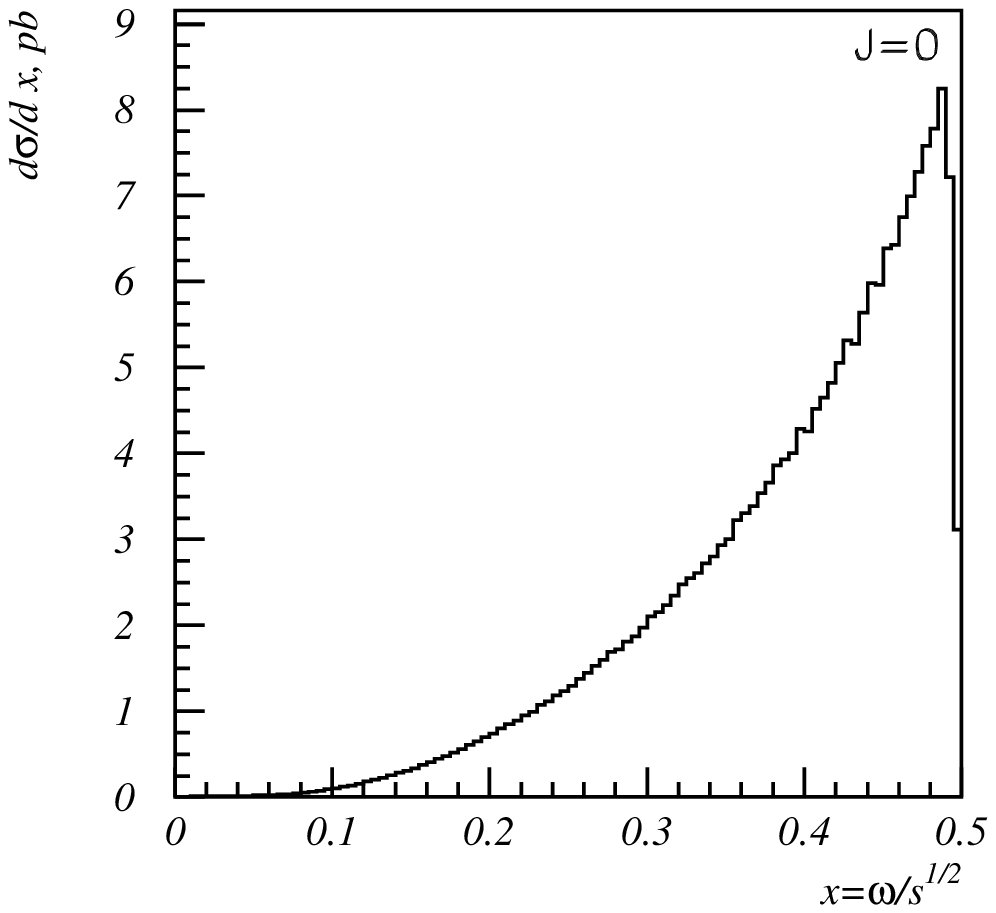}
\end{minipage}\hfill
\begin{minipage}[h]{.5\linewidth}
\centering
\includegraphics[width=\linewidth, height=3.5in, angle=0]{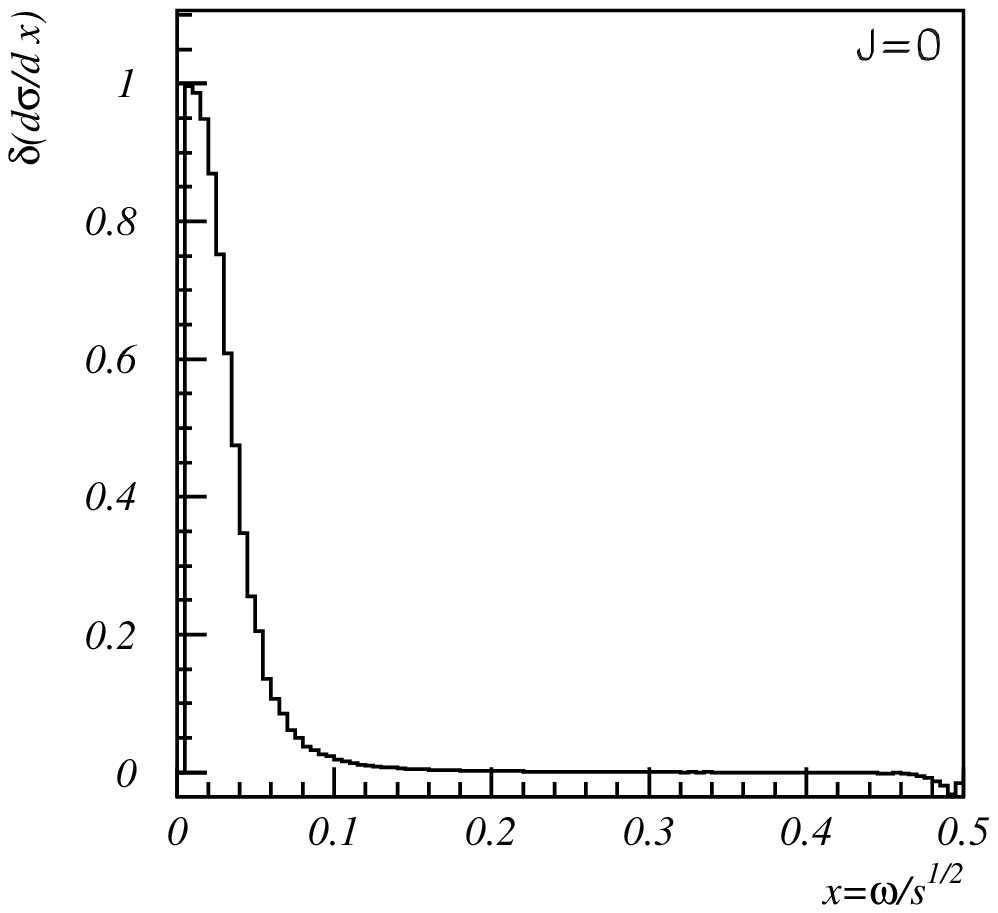}
\end{minipage}
\caption{
Energy spectra of final photon ($J\!=\!0$, $\sqrt {s} = 120 GeV$, $w_{cut}\!=\!1GeV$, $\E_{cut}\!=\!1GeV$, $\Theta_{cut}\!=\!7^o$, $\varphi_{cut}\!=\!3^o$).
}\label{f_en_1}
\end{figure}

Although the correction on the plots \ref{f_en_1} and \ref{f_hard_1} is high,
the differential cross section value at the point of peak is small
and does not affect the total cross section.

The calculations similar to ones made by us in ref. \cite{gg_llg_lumi} were recently performed in ref. \cite{kur}.
But it turned out that it is impossible to compare these results since the numerical results
or analysis of any kind are absent in ref. \cite{kur}.

Thus we can conclude that
massless analysis is applicable for $1\%$-level estimations of the $\gamma\gamma\to \mu\bar{\mu}\gamma$ process
except for some narrow regions of spectra of final particles.


\footnotesize

\newpage
$$ $$
\vspace{-1cm}

\vspace{-2cm}
$$ $$
\vspace{-1cm}

\begin{figure}[h!]
\leavevmode
\begin{minipage}[h]{.5\linewidth}
\centering
\includegraphics[width=\linewidth, height=3.5in, angle=0]{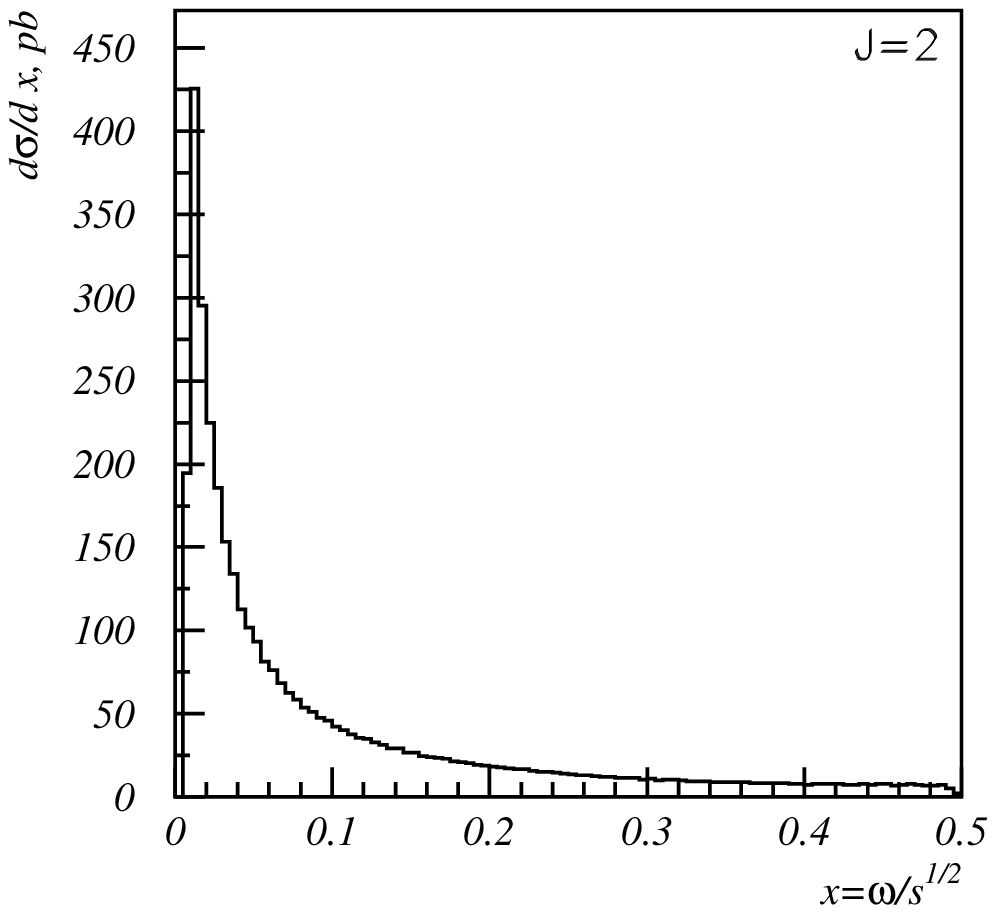}
\end{minipage}\hfill
\begin{minipage}[h]{.5\linewidth}
\centering
\includegraphics[width=\linewidth, height=3.5in, angle=0]{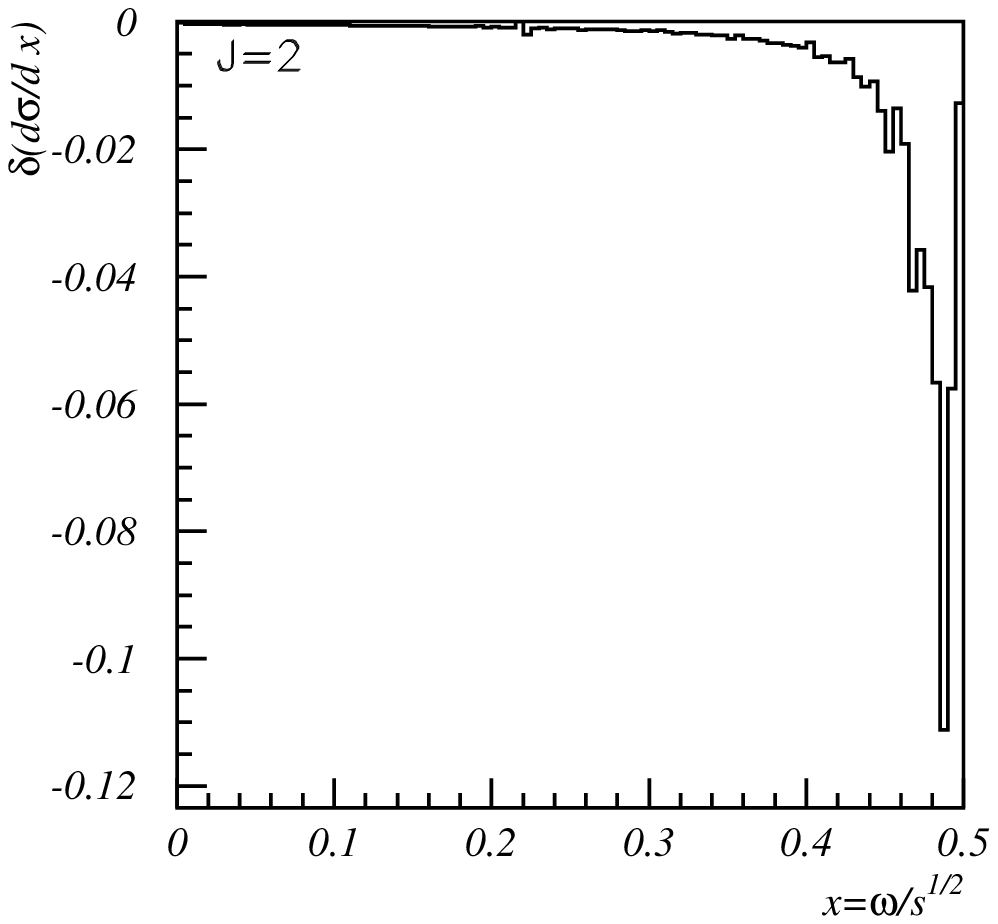}
\end{minipage}
\caption{
Energy spectra of final photon ($J\!=\!2$, $\sqrt {s} = 120 GeV$, $w_{cut}\!=\!1GeV$, $\E_{cut}\!=\!1GeV$, $\Theta_{cut}\!=\!7^o$, $\varphi_{cut}\!=\!3^o$).
}\label{f_en_2}
\end{figure}

$$ $$
\vspace{-1cm}

\newpage

\begin{figure}[h!]
\leavevmode
\begin{minipage}[h]{.33\linewidth}
\centering
\includegraphics[width=\linewidth, height=3.5in, angle=0]{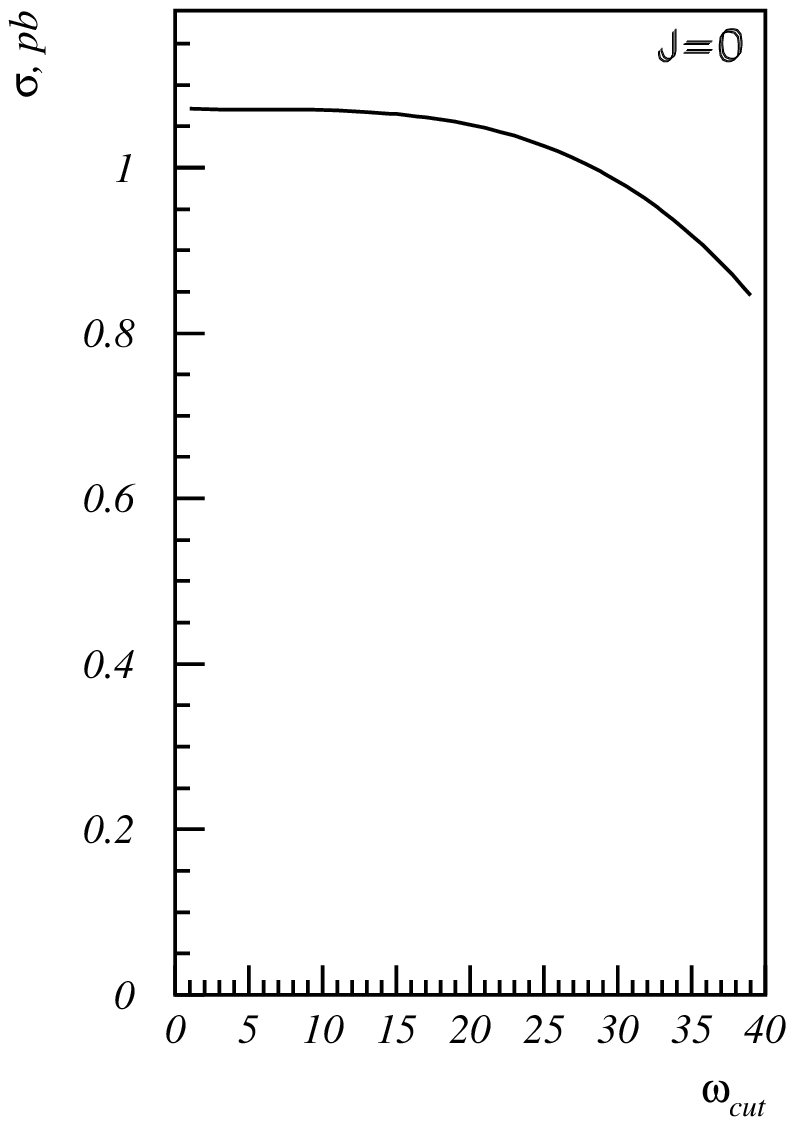}
\end{minipage}\hfill
\begin{minipage}[h]{.33\linewidth}
\centering
\includegraphics[width=\linewidth, height=3.5in, angle=0]{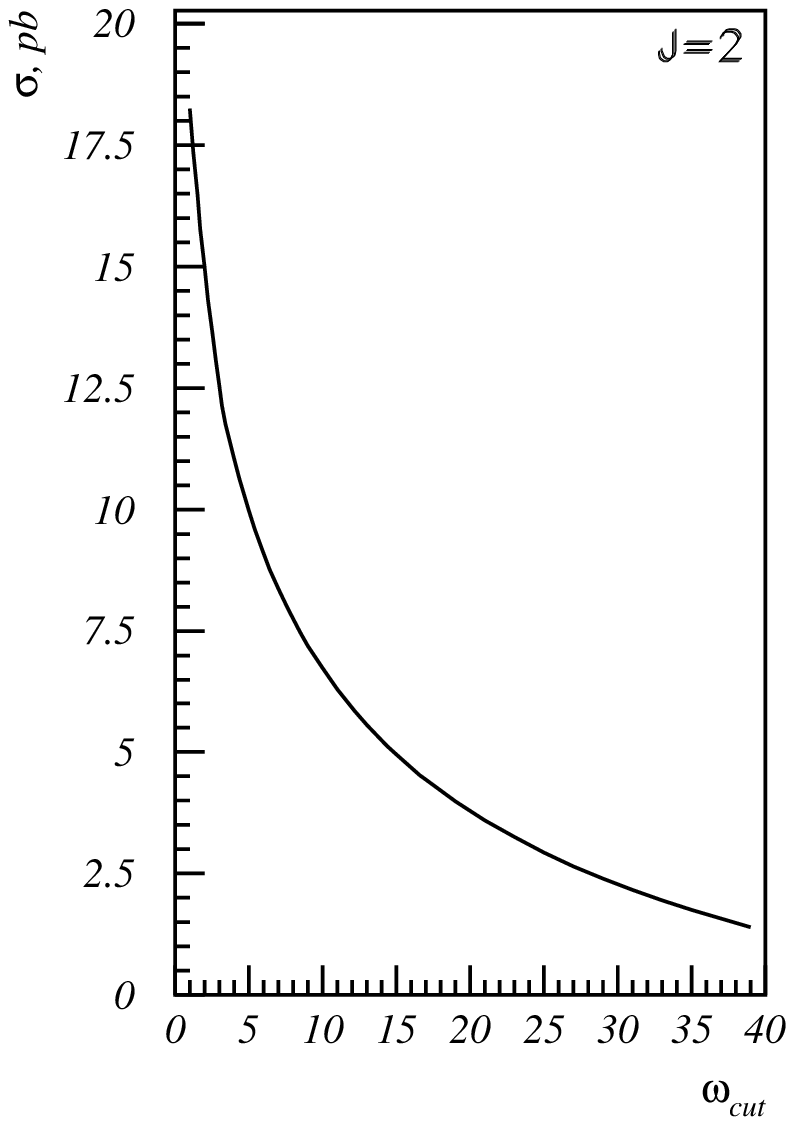}
\end{minipage}
\begin{minipage}[h]{.33\linewidth}
\centering
\includegraphics[width=\linewidth, height=3.5in, angle=0]{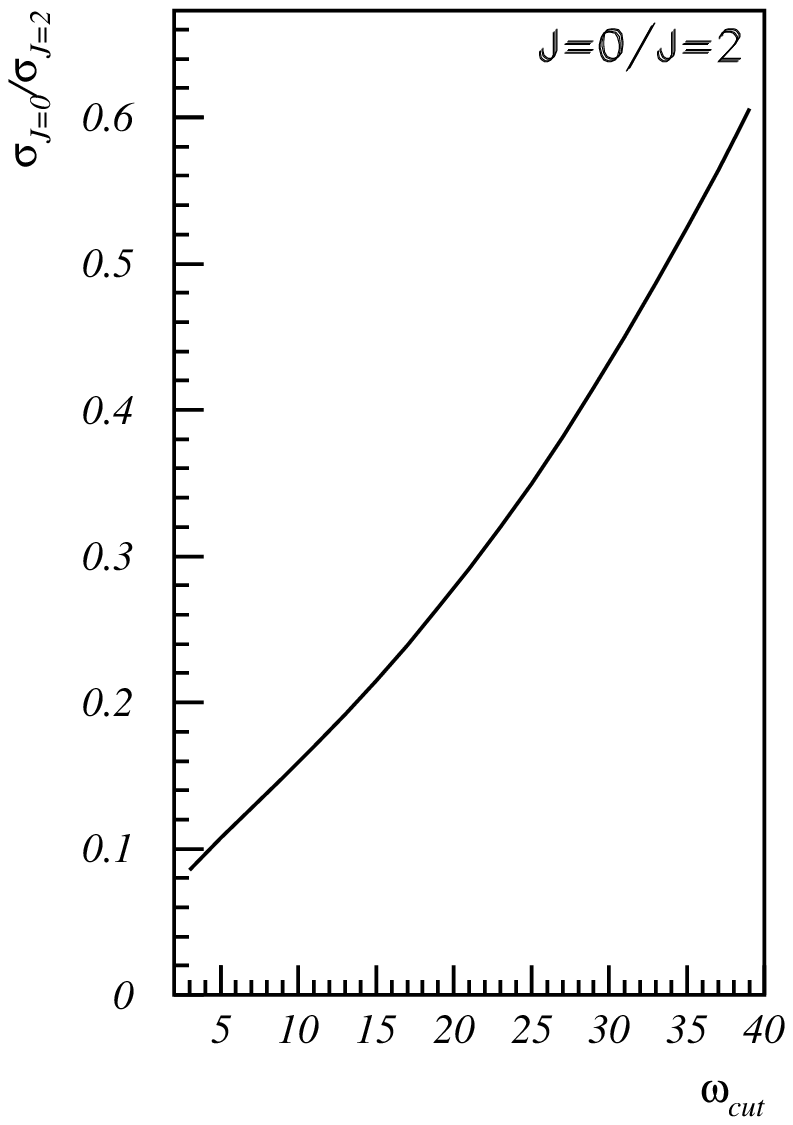}
\end{minipage}
\caption{
\small
Total cross section at $J\!=\!0$ and $J\!=\!2$ and their ratio (
$\sqrt {s} = 120 GeV$, $\E_{cut}\!=\!1GeV$, $\Theta_{cut}\!=\!7^o$, $\varphi_{cut}\!=\!3^o$)
at different cuts on the final-state photon energy.
}\label{f_tot_1}
\end{figure}

\vspace{-1cm}

\begin{figure}[h!]
\leavevmode
\begin{minipage}[h]{.33\linewidth}
\centering
\includegraphics[width=\linewidth, height=3.5in, angle=0]{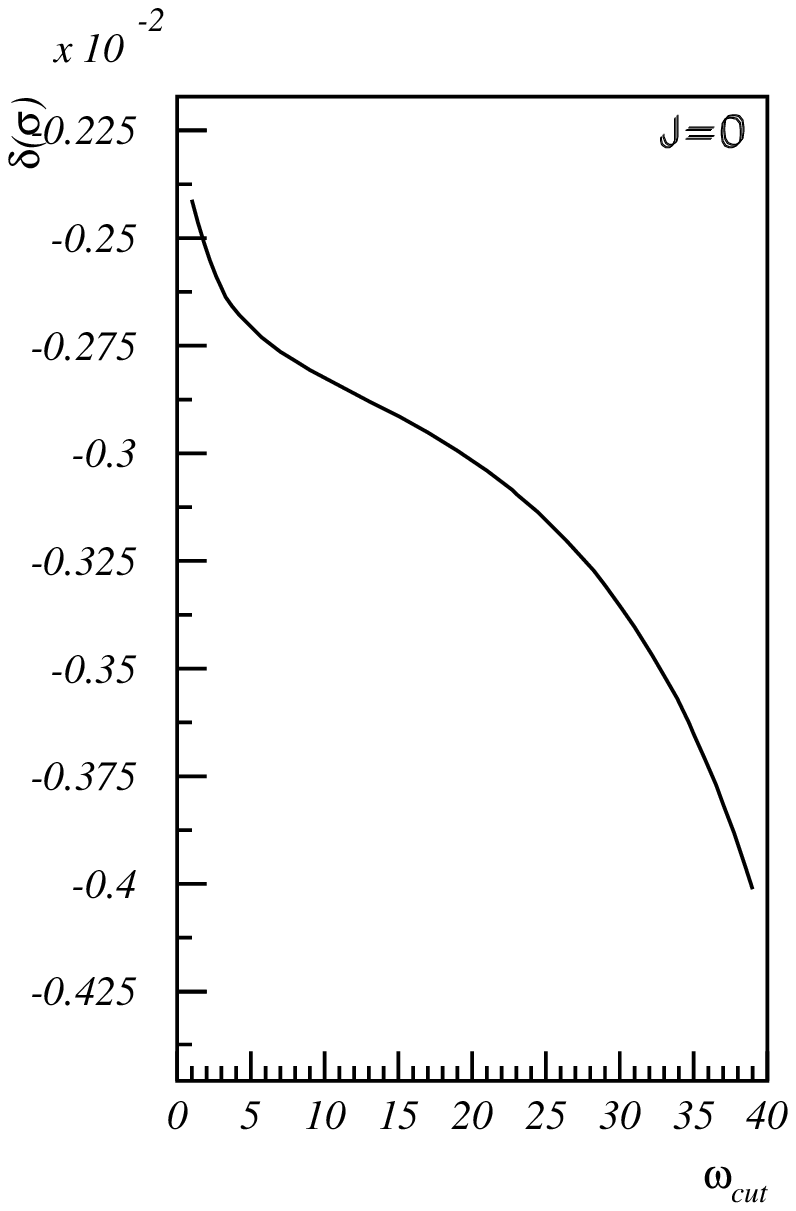}
\end{minipage}\hfill
\begin{minipage}[h]{.33\linewidth}
\centering
\includegraphics[width=\linewidth, height=3.5in, angle=0]{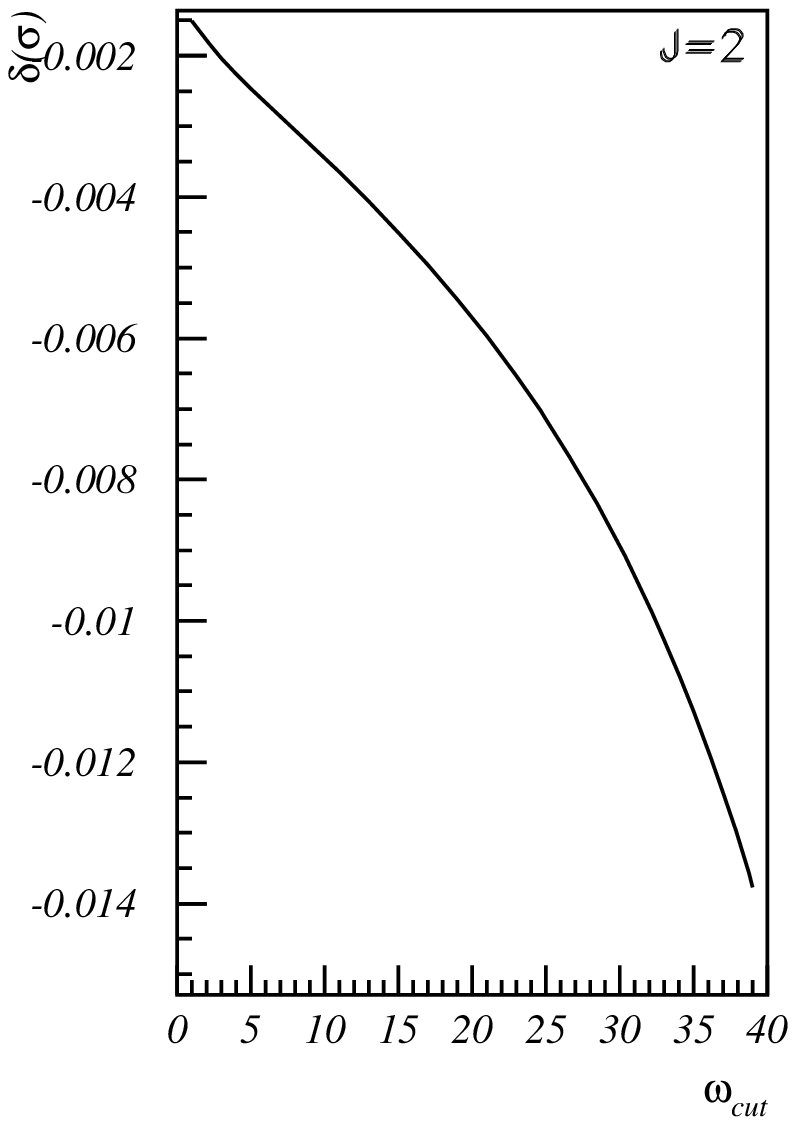}
\end{minipage}
\begin{minipage}[h]{.33\linewidth}
\centering
\includegraphics[width=\linewidth, height=3.5in, angle=0]{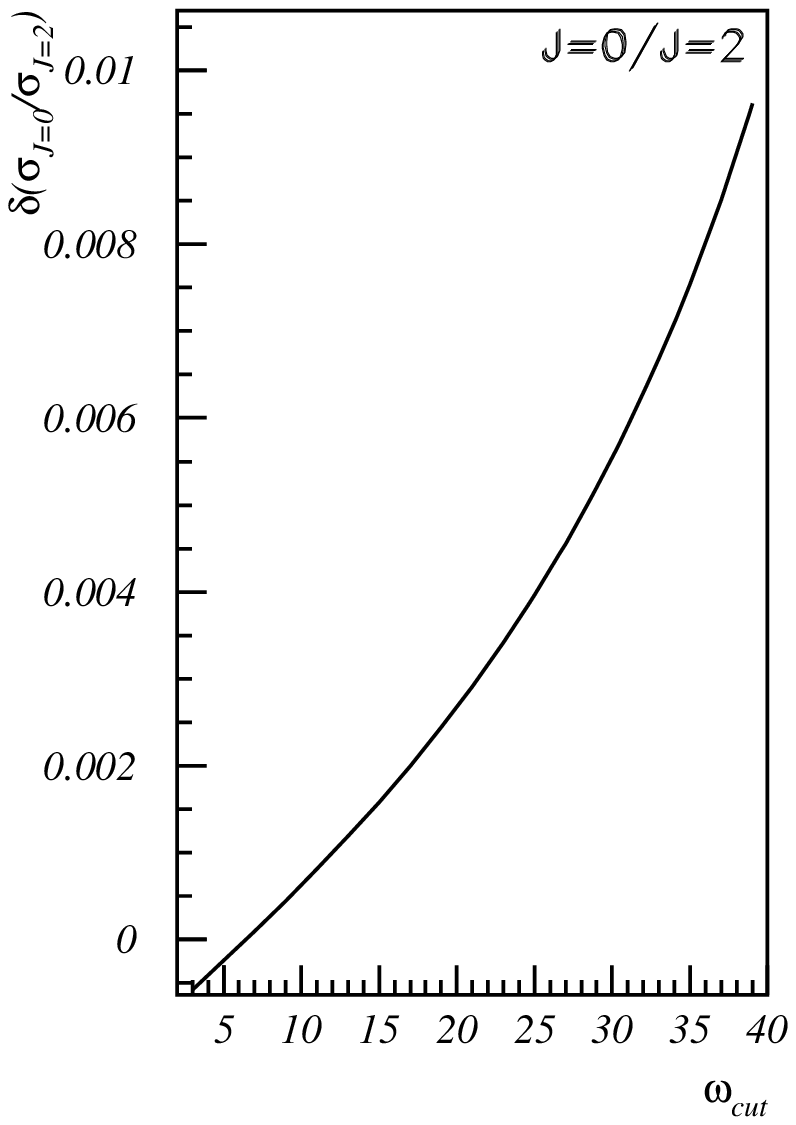}
\end{minipage}
\caption{
\small
Relative mass contributions to total cross sections
at different cuts on the final-state photon energy.
}\label{f_tot_delta_1}
\end{figure}

\newpage

\vspace{-1cm}
$$ $$
\vspace{-2cm}

\begin{figure}[h!]
\leavevmode
\begin{minipage}[h]{.5\linewidth}
\centering
\includegraphics[width=\linewidth, height=2.7in, angle=0]{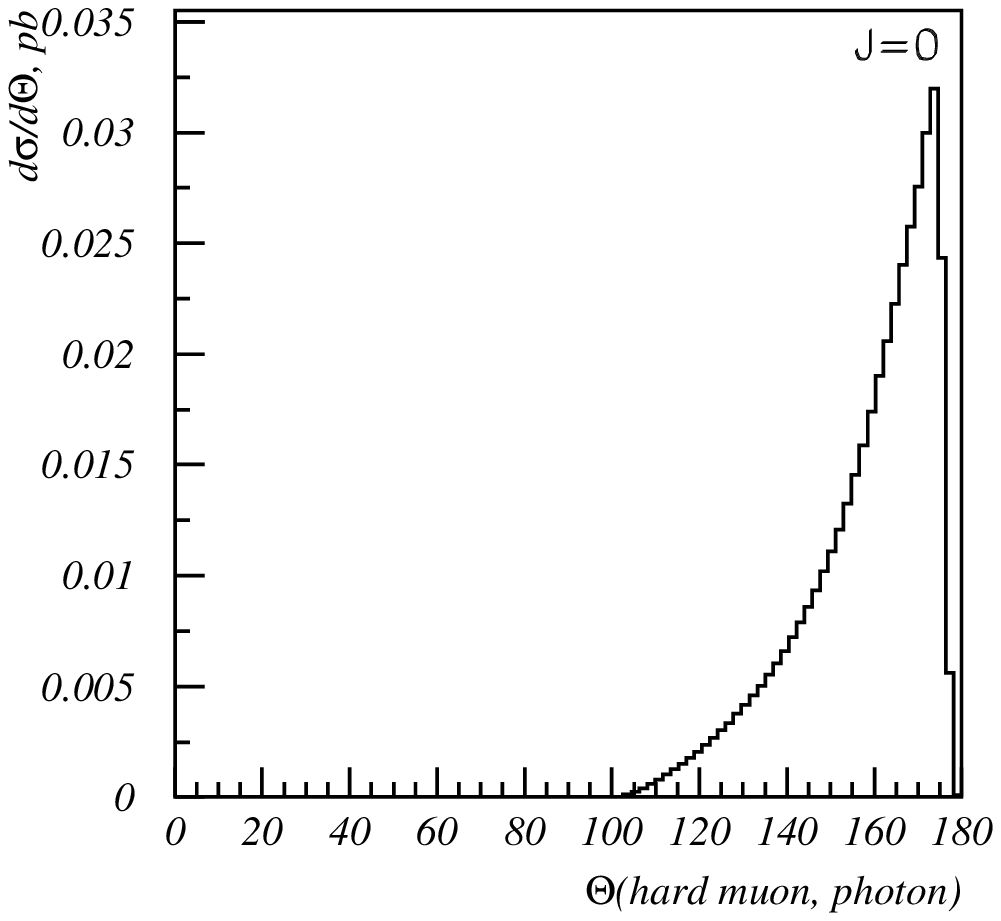}
\end{minipage}\hfill
\begin{minipage}[h]{.5\linewidth}
\centering
\includegraphics[width=\linewidth, height=2.7in, angle=0]{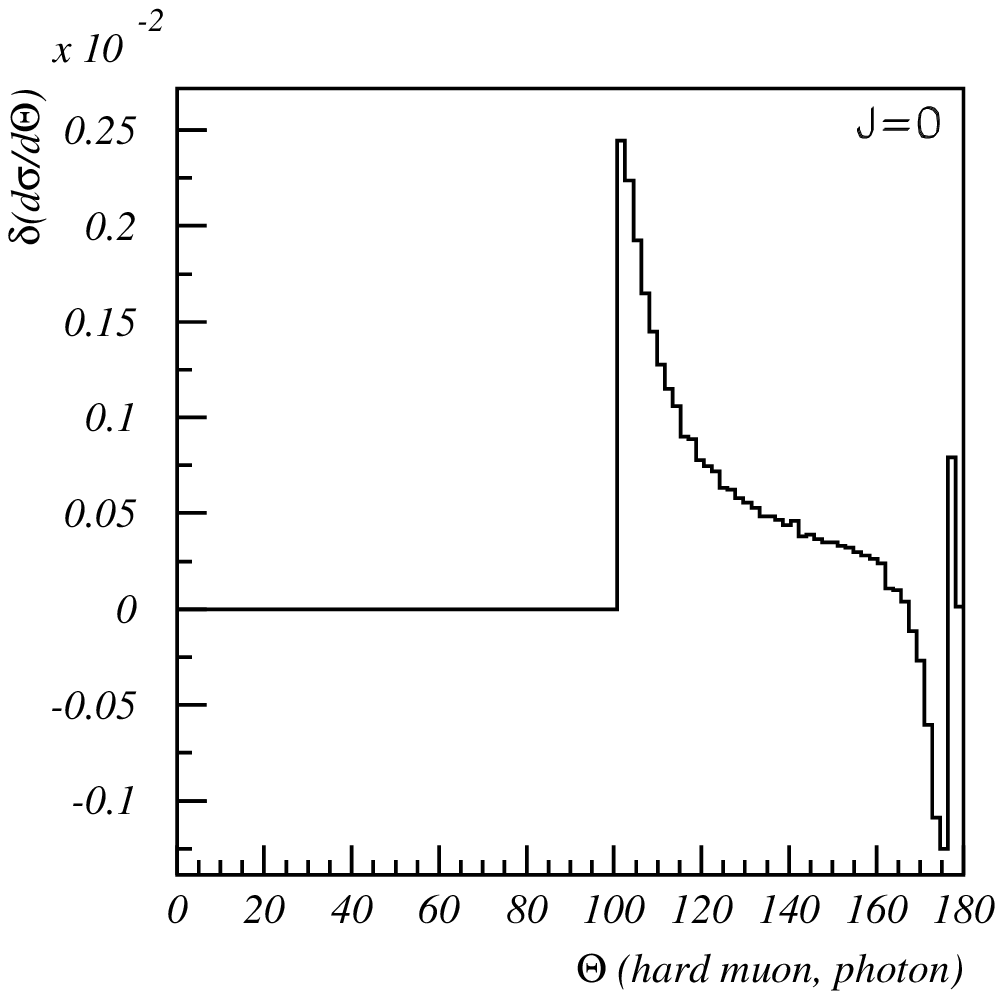}
\end{minipage}
\caption{
Angular spectrum of final particles,
$\sqrt {s} = 120 GeV$,
$w_{cut}\!=\!20GeV$, $\E_{cut}\!=\!5GeV$, $\Theta_{cut}\!=\!7^o$, $\varphi_{cut}\!=\!30^o$.
}\label{f_hard_2}
\end{figure}

\newpage

\begin{figure}[h!]
\centering
\begin{tabular}{|c|c|c|c|r|c|r|c|}
\hline
$\omega_{cut}, GeV$&
	$E_{f,cut}, GeV$&
		$\Theta_{cut}, {}^0$&
			$\varphi_{cut}, {}^0$&
				$\sigma {}_{J=0}, pb$&
					$\delta^{mass}_{J=0}$, \%&
						$\sigma {}_{J=2}, pb$&
							$\delta^{mass}_{J=0}$, \%\\

\hline
&&&&&&&\\

1&	1&	5&	3&	1.357&	-0.42&	21.065&	-0.15\\		
1&	1&	7&	3&	1.071&	-0.24&	18.248&	-0.15\\		
1&	1&	10&	3&	0.799&	-0.13&	15.288&	-0.15\\		
1&	1&	7&	5&	1.066&	-0.25&	15.789&	-0.07\\		
1&	1&	7&	10&	1.046&	-0.26&	12.472&	-0.03\\		
1&	1&	7&	30&	0.879&	-0.17&	7.157&	-0.02\\[3pt]	

10&	1&	7&	3&	1.069&	-0.28&	6.718&	-0.34\\		
10&	1&	7&	10&	1.044&	-0.23&	4.677&	-0.07\\		
10&	1&	7&	30&	0.878&	-0.19&	2.843&	-0.03\\		
10&	1&	5&	10&	1.312&	-0.48&	5.474&	-0.08\\		
10&	1&	10&	10&	0.784&	-0.16&	3.848&	-0.06\\[3pt]	

20&	1&	7&	3&	1.052&	-0.30&	3.789&	-0.57\\		
20&	1&	7&	10&	1.027&	-0.31&	2.690&	-0.12\\		
20&	5&	7&	10&	0.876&	-0.005&	2.561&	-0.04\\		
20&	5&	5&	30&	0.899&	-0.01&	1.905&	-0.03\\[5pt]	

\hline
\end{tabular}
\caption{
Total cross sections of $\gamma \gamma \to l^{+} l^{-} \gamma$ for different cuts
and relative mass contributions at $\sqrt {s} = 120 GeV$.
}
\end{figure}


\begin{thebibliography}{99}

\bibitem{tdr}

B. Badelek {\it et al.}, {\it TESLA Technical Design Report Part VI: The Photon Collider at TESLA},
DESY-01-011E, hep-ex/0108012.

\bibitem{gg_proposal}

I.F. Ginzburg, G.L. Kotkin, V.G. Serbo and V.I. Telnov, Nucl. Instr. Meth. {\bf 205} (1983) 47;

I.F. Ginzburg, G.L. Kotkin, S.L. Panfil, V.G. Serbo and V.I. Telnov, Nucl. Instr. Meth. {\bf 219} (1984) 5.

\bibitem{gg_2f}

A. Denner, S. Dittmaier, Eur. Phys. J. {\bf C9} (1999) 425, hep-ph/9812411.

\bibitem{gg_4f}

C. Carimalo, W. da Silva, F. Kapusta, Nucl. Phys. Proc. Suppl. {\bf 82} (2000) 391, hep-ph/9909339;

M. Moretti, Nucl. Phys. {\bf B484} (1997) 3, hep-ph/9604303.

\bibitem{gg_llg_lumi}

V. Makarenko, K. M\"onig, T. Shishkina, Eur. Phys. J. {\bf C30} d01 (2003) 011, LC-PHSM-2003-016, hep-ph/0306135.

\bibitem{gg_ffg}

T. V. Shishkina, V. V. Makarenko, hep-ph/0212409.

\bibitem{ha}

P. De Causmaecker et al, Phys. Lett. {\bf B105} (1981) 215;

F.A. Berends et al., Nucl. Phys. {\bf B206} (1982) 61.

\bibitem{mc}

S. Weinzierl, NIKHEF-00-012, hep-ph/0006269.

\bibitem{kur}

E. Bartos {\it et al.}, hep-ph/0308044.

\end{thebibliography}
\end{document}